\documentstyle[12pt]{article}

\begin{document}

\begin{flushright}
ITP-Budapest Report No. 508 \\
\vspace{1mm}
December 1994
\end{flushright}

\vspace{.2 in}
\begin{center}

{\Large \bf OBLIQUE RADIATIVE CORRECTIONS IN THE VECTOR CONDENSATE MODEL
    OF ELECTROWEAK INTERACTIONS}

\vspace{.1 in}

\large{G. Cynolter\footnote{E-mail: cyn@hal9000.elte.hu },
      E. Lendvai\footnote{E-mail: lendvai@hal9000.elte.hu} and
      G. P\'ocsik\footnote{E-mail: pocsik@ludens.elte.hu }  }
\vspace{2mm}

\large{     Institute for Theoretical Physics,
   Eotvos Lorand University, Budapest}

\end{center}

\begin{abstract}
{ Oblique radiative corrections are calculated to
the parameter $S$ in a version of the standard model where
the Higgs doublet is replaced by a doublet of vector bosons
and the gauge symmetry is broken dynamically. We show that
to each momentum scale there exists a domain of the masses
of charged and neutral vector bosons where $S$ is
compatible with the experiments. At a scale of 1 TeV this
requires vector boson masses of at least $m_0 \approx $ 400--550
GeV, $ m_+ \approx $ 200--350 GeV.}
\end{abstract}

\vspace{0.1 in}

The status of the standard model is very good, but it is
not clear whether the Higgs mechanism is responsible for
the origin of the mass. Therefore, alternative realizations
of the electroweak symmetry breaking are also important
[1].

Recently, we have proposed a model [2,3] of electroweak
symmetry breaking where the standard model scalar doublet
is replaced by a doublet of vector fields,
\begin{equation}
 B_{ \mu } = \pmatrix{ B_{ \mu }^{(+)} \cr
 B_{ \mu }^{(0)} \cr} ,
\end{equation}
and $B_{ \mu }^{(0)}$ forms a condensate
\begin{equation}
 \left \langle B_{ \mu }^{(0)+} B_{ \nu }^{(0)} \right \rangle_0
  = g_{ \mu \nu } d , \qquad d \not = 0 .
\end{equation}
$ B_{ \mu } $ is coupled to the $SU(2) \times U(1) $ gauge fields
and to itself by a gauge invariant Lagrangian
\begin{equation}
L=-{1 \over 2} \left( D_{\mu}B_{\nu}-D_{\nu}B_{\mu}
\right)^+ \left( D^{\mu}B^{\nu}-D^{\nu}B^{\mu} \right)
-\lambda (B_{\nu}^+ B^{\nu} )^2, \lambda >0,
\end{equation}
with $D_{\mu}$ the usual covariant derivative. Fermion and
gauge field couplings are standard.

Breaking the symmetry in (3) by (2) gives rise to the usual
mass terms for the gauge fields and makes also the B
particles massive. $\sqrt{-6d} \simeq 246 $GeV plays the role
of the vacuum expectation value of the scalar field [2] and
$\rho_{tree}=1$.

Fermion masses are considered as manifestations of the
condensation (2) and derived from the Lagrangian

$$ g_{ij}^u \left ( \overline{\Psi}_{Li} B_{\nu}^C u_{Rj} \right )
 B_{\nu}^{(0)}+
 g_{ij}^d \left ( \overline{\Psi}_{Li} B_{\nu} d_{Rj} \right )
 B_{\nu}^{(0)+}+h.c. ,$$
\begin{equation}
\Psi_{Li}=\pmatrix{u_i \cr d_i \cr }_L
B_{\nu}^C=\pmatrix{ B_{ \nu }^{(0)+} \cr - B_{ \nu}^{(+)+} \cr}
\end{equation}
(and similarly for leptons) [3]. (4) allows the Kobayashi--Maskawa
mechanism, too.  (3) and (4) fix the interactions
of B particles. Producing B particles in $e^+e^-$
annihilation has been investigated in ref. [2].

The model can be considered as a low energy effective model valid up to
a scale $\Lambda, \  \Lambda \approx 2.6 $TeV. We have shown [3] that
oblique radiative corrections due to B--loops give arbitrarily small
contributions to the $\rho$ parameter if $B^{+,0}$ masses
are suitably chosen, and $ \Lambda $ remains unrestricted.

In the present paper one-loop oblique radiative corrections
are calculated to the parameter S, one of the three
parameters [4] constrained by precision experiments. It is
shown that the B particles must be heavy and to each
$\Lambda $ there exists a domain of $B^{+,0}$ masses where
the model remains valid. For B masses which are very large
compared to a fixed $\Lambda $, B--loops contribute
negligibly to S. At a scale of 1 TeV the threshold is about
$m_+ \approx $ 200--350 GeV, $m_0 \approx $ 400--550 GeV.

The parameter S [4] defined by
$$ \alpha S=4 e^2 \big( \overline{ \Pi}'_{ZZ}(0) - (c^2\!-\!s^2)
\overline{\Pi}'_{ZA}(0)- s^2 c^2 \overline{\Pi}'_{AA}(0)
\big ), $$
\begin{equation}
\overline{\Pi}'_{ik}(0) = \frac{d}{dq^2}
\overline{\Pi}'_{ik}(q^2) \vert_{q^2=0}
\end{equation}
with $c=\cos \theta_w, \  s=\sin \theta_w $ is
calculated in one B--loop order.
$\overline{\Pi}'_{ik}(q^2)$ is expressed by the $g_{\mu
\nu}$ terms of the vacuum polarization contributions $
\Pi_{ik}(q^2) $ due to B--loops as

\begin{equation}
\Pi_{ZZ}=\frac{e^2}{s^2 c^2} \overline{ \Pi}_{ZZ}, \:
\Pi_{ZA}=\frac{e^2}{s c} \overline{ \Pi}_{ZA}, \:
\Pi_{AA}={e^2} \overline{ \Pi}_{AA}.
\end{equation}

To one B--loop order only trilinear interactions of B are
essential, from (3) these are as follows
$$ L\left( B^{0} \right)={ig \over 2c } \partial^{\mu} { B^{(0)
\nu }}^+ \left( Z_{ \mu} B_{\nu}^{(0)} - Z_{\nu} B^{(0)}_{\mu}
\right) + h.c. , $$
\begin{equation}
L\left( B^+ {B^+}^+ Z \right)=-(c^2-s^2) \cdot L\
\left(B^0 \rightarrow B^+ \right),
\end{equation}
$$ L\left(B^+B^-A\right)=ie \partial^{\mu} { B^{(+)
\nu }}^+ \left( A_{ \mu} B_{\nu}^{(+)} - A_{\nu} B^{(+)}_{\mu}
\right) + h.c. . $$

In a renormalizable model S is finite. In the model based
on (3), however, cancellations of $\Lambda$ powers are not
perfect due to nonrenormalizability.  We get for the scaled
vacuum polarizations $\overline{\Pi}'_{ik}(0)$,

$$\overline{\Pi}'_{ZZ}(0)=\frac{1}{128 \pi^2}
\left( f \left( \frac{\Lambda^2}{m_0^2} \right) +(c^2-s^2)^2
f\left( \frac{\Lambda^2}{m_+^2} \right)  \right),$$
\begin{equation}
 \overline{\Pi}'_{ZA}(0)=-\frac{c^2-s^2}{64 \pi^2}
f\left( \frac{\Lambda^2}{m_+^2} \right) ,
\end{equation}
$$ \overline{\Pi}'_{AA}(0)=\frac{1}{32 \pi^2}
f\left( \frac{\Lambda^2}{m_+^2} \right) ,  $$

where $m_+ (m_0)$ is the physical mass of $B^+ (B^0) $, and
\begin{equation}
f(x)=5x-16 \ln(1+x)+14-\frac{17}{1+x}+\frac{3}{(1+x)^2}.
\end{equation}

For increasing m $f \left( \frac{\Lambda^2}{m^2} \right) $
decreases through f=0 at $ m=1.90 \Lambda $, and at
$\frac{\Lambda}{m} \rightarrow 0 \quad f \rightarrow 0 $ from
below.

Substituting (8) into (5) we obtain

\begin{equation}
S=\frac{1}{8 \pi} \ \left[ f \left( \frac{\Lambda^2}{m_0^2} \right)
+(3(c^2\!-\!s^2)^2 -4 s^2 c^2 ) f \left( \frac{\Lambda^2}{m_+^2}
\right) \right].
\end{equation}

The coefficient of $f\left( \frac{\Lambda^2}{m_+^2}
\right) $ is 0.158 for $s^2=0.231$.

An analysis of precision experiments shows that
$S_{new}<0.09 (0.23) $ at 90 (95)\% C.L. [5] for
$m_H^{ref}=300 $GeV and assuming $m_t=174 $GeV (CDF value).
Requiring $S_{new} \geq 0 $, the corresponding constraints
are $S_{new} <0.38 (0.46) $ [5]. Since a Higgs of 300 GeV
is absent in the present model, its contribution, 0.063,
must be removed. In this way for the contribution of B we
have $S<0.15 (0.29)$ at 90 (95)\% C.L.  For $m_+, \  m_0
\geq 1.90 \Lambda$ this is fulfilled, in particular, $S
\rightarrow 0 $ for $\frac{\Lambda}{m_{+,0}} \rightarrow 0
$. One can fulfil the experimental constraints also for a
mass(es) lower than $1.90 \Lambda $. This is shown in
Fig.1. for $\Lambda=1 $TeV, where the curves are coming
from the 90\% and 95\% C.L. limits on S for $S_{new} \geq 0 $
(lower two curves) and for $S\frac{>}{<} 0$ (upper two
curves), respectively. Excluded regions are below the
curves. It follows that the B particles are heavy; at
$\Lambda=1 $TeV the threshold is about $m_0 \approx$ 400--550
GeV and $m_+ \approx $ 200--350 GeV. Since S is invariant
multiplying $\Lambda, m_0, m_+ $ by a common factor, Fig.1.
determines the allowed regions for scales different from  1
TeV. Higher $\Lambda$ attracts higher minimum masses.
Recently, we have shown [6] that B particles could be seen
at high energy linear $e^+e^-$ colliders up to masses of
several hundred GeV's.

In conclusion, in the vector condensate model, to each momentum
scale there exists a range of B boson masses where S is
compatible with the experimental constraints. Further
restrictions are imposed on $m_{+,0}$ by taking into account the
results [3] on the parameter T. While for a fixed $\Lambda$ and
$m_0$ a large $m_+$ range exists from S (see Fig.1.), this is
tightened by T. For example, at $\Lambda= 1$TeV, $m_0=$400(600)
GeV, the $m_+$ range allowed by S, T is $m_+=$630--636(846--868)
GeV. For higher $\Lambda $ the allowed $m_+$ region shrinks at
the same $m_0$.

In general, $\Lambda$ remains unrestricted and suitable, heavy
$B^{+,0}$ provide small radiative corrections.

\vskip 20pt
{\bf Acknowledgment}

This work is supported in part by OTKA I/3, No. 2190.

\vspace{1 in}

{\bf REFERENCES}
\newcounter{bib}
\begin{list}%
{ [{\arabic{bib}}]}{\usecounter{bib} \setlength{\rightmargin}{\leftmargin} }

\item {  Proc. of LHC Workshop, Aachen 1990, CERN 90-10, ed.
G.Jarlskog and D.Rein, Vol.2, p.757; M.Lindner,
Int.J.Mod.Phys. A8, (1993) 2167; R.Casalbuoni, S.De Curtis
and M.Grazzini, Phys.Lett.B 317 (1993) 151.}

\item {G.P\'ocsik, E.Lendvai and G.Cynolter, Acta Phys. Polonica
B 24 (1993) 1495.}

{\item G.Gynolter, E.Lendvai and G.P\'ocsik, Mod. Phys. Lett. A 9 (1994) 1701.}

{\item D.Kennedy and B.W.Lynn, Nucl. Phys. B 322 (1989) 1;
     M.E.Peskin and T.Takeuchi, Phys.Rev.Lett. 65 (1990) 964;
     G.Altarelli and R.Barbieri, Phys.Lett. B253 (1990) 161.}

{\item P.Langacker and J.Erler, in Review of Particle
Properties, Phys.Rev. D 50 (1994) Nu.3, 1312.}

{\item G.Cynolter, E.Lendvai and G.P\'ocsik,
ITP-Budapest Report, No.507, l994.}

\end{list}

\newpage

{\bf FIGURE CAPTION}

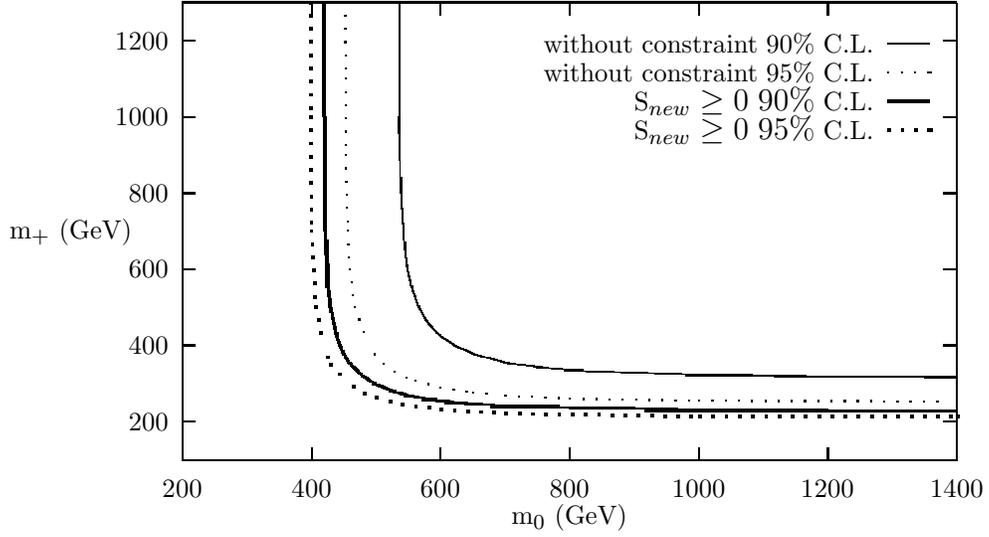
Fig. 1.  $m_+$ vs. $m_0$ at  $\Lambda=1 $TeV. Lower (upper)
two curves embody the 90\% and 95\% C.L. on S for
$S_{new}\geq 0 $ ($S_{new}\frac{>}{<} 0$). Allowed regions
are above the curves.

\setlength{\unitlength}{0.240900pt}
\ifx\plotpoint\undefined\newsavebox{\plotpoint}\fi
\sbox{\plotpoint}{\rule[-0.200pt]{0.400pt}{0.400pt}}%
\begin{picture}(1500,900)(0,0)
\font\gnuplot=cmr10 at 10pt
\gnuplot
\sbox{\plotpoint}{\rule[-0.200pt]{0.400pt}{0.400pt}}%
\put(220.0,173.0){\rule[-0.200pt]{4.818pt}{0.400pt}}
\put(198,173){\makebox(0,0)[r]{200}}
\put(1416.0,173.0){\rule[-0.200pt]{4.818pt}{0.400pt}}
\put(220.0,293.0){\rule[-0.200pt]{4.818pt}{0.400pt}}
\put(198,293){\makebox(0,0)[r]{400}}
\put(1416.0,293.0){\rule[-0.200pt]{4.818pt}{0.400pt}}
\put(220.0,413.0){\rule[-0.200pt]{4.818pt}{0.400pt}}
\put(198,413){\makebox(0,0)[r]{600}}
\put(1416.0,413.0){\rule[-0.200pt]{4.818pt}{0.400pt}}
\put(220.0,532.0){\rule[-0.200pt]{4.818pt}{0.400pt}}
\put(198,532){\makebox(0,0)[r]{800}}
\put(1416.0,532.0){\rule[-0.200pt]{4.818pt}{0.400pt}}
\put(220.0,652.0){\rule[-0.200pt]{4.818pt}{0.400pt}}
\put(198,652){\makebox(0,0)[r]{1000}}
\put(1416.0,652.0){\rule[-0.200pt]{4.818pt}{0.400pt}}
\put(220.0,772.0){\rule[-0.200pt]{4.818pt}{0.400pt}}
\put(198,772){\makebox(0,0)[r]{1200}}
\put(1416.0,772.0){\rule[-0.200pt]{4.818pt}{0.400pt}}
\put(220.0,113.0){\rule[-0.200pt]{0.400pt}{4.818pt}}
\put(220,68){\makebox(0,0){200}}
\put(220.0,812.0){\rule[-0.200pt]{0.400pt}{4.818pt}}
\put(423.0,113.0){\rule[-0.200pt]{0.400pt}{4.818pt}}
\put(423,68){\makebox(0,0){400}}
\put(423.0,812.0){\rule[-0.200pt]{0.400pt}{4.818pt}}
\put(625.0,113.0){\rule[-0.200pt]{0.400pt}{4.818pt}}
\put(625,68){\makebox(0,0){600}}
\put(625.0,812.0){\rule[-0.200pt]{0.400pt}{4.818pt}}
\put(828.0,113.0){\rule[-0.200pt]{0.400pt}{4.818pt}}
\put(828,68){\makebox(0,0){800}}
\put(828.0,812.0){\rule[-0.200pt]{0.400pt}{4.818pt}}
\put(1031.0,113.0){\rule[-0.200pt]{0.400pt}{4.818pt}}
\put(1031,68){\makebox(0,0){1000}}
\put(1031.0,812.0){\rule[-0.200pt]{0.400pt}{4.818pt}}
\put(1233.0,113.0){\rule[-0.200pt]{0.400pt}{4.818pt}}
\put(1233,68){\makebox(0,0){1200}}
\put(1233.0,812.0){\rule[-0.200pt]{0.400pt}{4.818pt}}
\put(1436.0,113.0){\rule[-0.200pt]{0.400pt}{4.818pt}}
\put(1436,68){\makebox(0,0){1400}}
\put(1436.0,812.0){\rule[-0.200pt]{0.400pt}{4.818pt}}
\put(220.0,113.0){\rule[-0.200pt]{292.934pt}{0.400pt}}
\put(1436.0,113.0){\rule[-0.200pt]{0.400pt}{173.207pt}}
\put(220.0,832.0){\rule[-0.200pt]{292.934pt}{0.400pt}}
\put(45,472){\makebox(0,0){m$_+$ (GeV)}}
\put(828,23){\makebox(0,0){m$_0$ (GeV)}}
\put(828,877){\makebox(0,0){Fig.1.}}
\put(220.0,113.0){\rule[-0.200pt]{0.400pt}{173.207pt}}
\put(1306,767){\makebox(0,0)[r]{without constraint 90\% C.L.}}
\put(1328.0,767.0){\rule[-0.200pt]{15.899pt}{0.400pt}}
\put(560,832){\usebox{\plotpoint}}
\multiput(560.60,601.77)(0.468,-17.443){5}{\rule{0.113pt}{12.100pt}}
\multiput(559.17,626.89)(4.000,-94.886){2}{\rule{0.400pt}{6.050pt}}
\multiput(564.59,506.96)(0.482,-7.994){9}{\rule{0.116pt}{6.033pt}}
\multiput(563.17,519.48)(6.000,-76.478){2}{\rule{0.400pt}{3.017pt}}
\multiput(570.60,430.13)(0.468,-4.283){5}{\rule{0.113pt}{3.100pt}}
\multiput(569.17,436.57)(4.000,-23.566){2}{\rule{0.400pt}{1.550pt}}
\multiput(574.59,405.47)(0.485,-2.247){11}{\rule{0.117pt}{1.814pt}}
\multiput(573.17,409.23)(7.000,-26.234){2}{\rule{0.400pt}{0.907pt}}
\multiput(581.58,378.06)(0.492,-1.392){19}{\rule{0.118pt}{1.191pt}}
\multiput(580.17,380.53)(11.000,-27.528){2}{\rule{0.400pt}{0.595pt}}
\multiput(592.58,349.65)(0.493,-0.893){23}{\rule{0.119pt}{0.808pt}}
\multiput(591.17,351.32)(13.000,-21.324){2}{\rule{0.400pt}{0.404pt}}
\multiput(605.58,327.76)(0.496,-0.549){37}{\rule{0.119pt}{0.540pt}}
\multiput(604.17,328.88)(20.000,-20.879){2}{\rule{0.400pt}{0.270pt}}
\multiput(625.00,306.92)(0.754,-0.494){25}{\rule{0.700pt}{0.119pt}}
\multiput(625.00,307.17)(19.547,-14.000){2}{\rule{0.350pt}{0.400pt}}
\multiput(646.00,292.92)(1.084,-0.494){25}{\rule{0.957pt}{0.119pt}}
\multiput(646.00,293.17)(28.013,-14.000){2}{\rule{0.479pt}{0.400pt}}
\multiput(676.00,278.92)(1.856,-0.494){25}{\rule{1.557pt}{0.119pt}}
\multiput(676.00,279.17)(47.768,-14.000){2}{\rule{0.779pt}{0.400pt}}
\multiput(727.00,264.93)(3.772,-0.485){11}{\rule{2.957pt}{0.117pt}}
\multiput(727.00,265.17)(43.862,-7.000){2}{\rule{1.479pt}{0.400pt}}
\multiput(777.00,257.93)(5.608,-0.477){7}{\rule{4.180pt}{0.115pt}}
\multiput(777.00,258.17)(42.324,-5.000){2}{\rule{2.090pt}{0.400pt}}
\multiput(828.00,252.93)(13.371,-0.488){13}{\rule{10.250pt}{0.117pt}}
\multiput(828.00,253.17)(181.726,-8.000){2}{\rule{5.125pt}{0.400pt}}
\put(1031,244.17){\rule{40.500pt}{0.400pt}}
\multiput(1031.00,245.17)(117.940,-2.000){2}{\rule{20.250pt}{0.400pt}}
\put(560.0,652.0){\rule[-0.200pt]{0.400pt}{43.362pt}}
\put(1335,242.67){\rule{24.331pt}{0.400pt}}
\multiput(1335.00,243.17)(50.500,-1.000){2}{\rule{12.165pt}{0.400pt}}
\put(1233.0,244.0){\rule[-0.200pt]{24.572pt}{0.400pt}}
\put(1306,722){\makebox(0,0)[r]{without constraint 95\% C.L.}}
\multiput(1328,722)(20.756,0.000){4}{\usebox{\plotpoint}}
\put(1394,722){\usebox{\plotpoint}}
\put(475,832){\usebox{\plotpoint}}
\multiput(475,832)(0.173,-20.755){6}{\usebox{\plotpoint}}
\multiput(476,712)(0.000,-20.756){6}{\usebox{\plotpoint}}
\multiput(476,592)(0.523,-20.749){6}{\usebox{\plotpoint}}
\multiput(479,473)(1.036,-20.730){3}{\usebox{\plotpoint}}
\multiput(482,413)(2.743,-20.573){3}{\usebox{\plotpoint}}
\multiput(490,353)(6.857,-19.590){3}{\usebox{\plotpoint}}
\put(521.05,279.08){\usebox{\plotpoint}}
\put(535.69,264.48){\usebox{\plotpoint}}
\put(552.31,252.25){\usebox{\plotpoint}}
\put(570.70,242.72){\usebox{\plotpoint}}
\multiput(585,237)(20.136,-5.034){2}{\usebox{\plotpoint}}
\multiput(625,227)(20.505,-3.216){3}{\usebox{\plotpoint}}
\multiput(676,219)(20.656,-2.025){2}{\usebox{\plotpoint}}
\multiput(727,214)(20.730,-1.026){5}{\usebox{\plotpoint}}
\multiput(828,209)(20.753,-0.307){10}{\usebox{\plotpoint}}
\multiput(1031,206)(20.755,-0.051){19}{\usebox{\plotpoint}}
\put(1436,205){\usebox{\plotpoint}}
\sbox{\plotpoint}{\rule[-0.400pt]{0.800pt}{0.800pt}}%
\put(1306,677){\makebox(0,0)[r]{S$_{new} \geq 0 \  90\%$ C.L.}}
\put(1328.0,677.0){\rule[-0.400pt]{15.899pt}{0.800pt}}
\put(442,832){\usebox{\plotpoint}}
\put(440.84,652){\rule{0.800pt}{43.362pt}}
\multiput(440.34,742.00)(1.000,-90.000){2}{\rule{0.800pt}{21.681pt}}
\put(442.34,473){\rule{0.800pt}{43.121pt}}
\multiput(441.34,562.50)(2.000,-89.500){2}{\rule{0.800pt}{21.561pt}}
\put(445.34,383){\rule{0.800pt}{18.200pt}}
\multiput(443.34,435.22)(4.000,-52.225){2}{\rule{0.800pt}{9.100pt}}
\put(449.34,353){\rule{0.800pt}{6.200pt}}
\multiput(447.34,370.13)(4.000,-17.132){2}{\rule{0.800pt}{3.100pt}}
\multiput(454.40,336.23)(0.514,-2.602){13}{\rule{0.124pt}{4.040pt}}
\multiput(451.34,344.61)(10.000,-39.615){2}{\rule{0.800pt}{2.020pt}}
\multiput(464.40,296.20)(0.514,-1.267){13}{\rule{0.124pt}{2.120pt}}
\multiput(461.34,300.60)(10.000,-19.600){2}{\rule{0.800pt}{1.060pt}}
\multiput(474.40,274.74)(0.512,-0.838){15}{\rule{0.123pt}{1.509pt}}
\multiput(471.34,277.87)(11.000,-14.868){2}{\rule{0.800pt}{0.755pt}}
\multiput(484.00,261.09)(0.524,-0.506){29}{\rule{1.044pt}{0.122pt}}
\multiput(484.00,261.34)(16.832,-18.000){2}{\rule{0.522pt}{0.800pt}}
\multiput(503.00,243.08)(0.762,-0.511){17}{\rule{1.400pt}{0.123pt}}
\multiput(503.00,243.34)(15.094,-12.000){2}{\rule{0.700pt}{0.800pt}}
\multiput(521.00,231.08)(1.259,-0.511){17}{\rule{2.133pt}{0.123pt}}
\multiput(521.00,231.34)(24.572,-12.000){2}{\rule{1.067pt}{0.800pt}}
\multiput(550.00,219.08)(2.052,-0.526){7}{\rule{3.057pt}{0.127pt}}
\multiput(550.00,219.34)(18.655,-7.000){2}{\rule{1.529pt}{0.800pt}}
\multiput(575.00,212.07)(3.141,-0.536){5}{\rule{4.200pt}{0.129pt}}
\multiput(575.00,212.34)(21.283,-6.000){2}{\rule{2.100pt}{0.800pt}}
\multiput(605.00,206.06)(6.469,-0.560){3}{\rule{6.760pt}{0.135pt}}
\multiput(605.00,206.34)(26.969,-5.000){2}{\rule{3.380pt}{0.800pt}}
\multiput(646.00,201.07)(8.834,-0.536){5}{\rule{11.000pt}{0.129pt}}
\multiput(646.00,201.34)(58.169,-6.000){2}{\rule{5.500pt}{0.800pt}}
\multiput(727.00,195.07)(33.728,-0.536){5}{\rule{40.733pt}{0.129pt}}
\multiput(727.00,195.34)(219.456,-6.000){2}{\rule{20.367pt}{0.800pt}}
\put(1031,188.84){\rule{97.564pt}{0.800pt}}
\multiput(1031.00,189.34)(202.500,-1.000){2}{\rule{48.782pt}{0.800pt}}
\sbox{\plotpoint}{\rule[-0.500pt]{1.000pt}{1.000pt}}%
\put(1306,632){\makebox(0,0)[r]{S$_{new} \geq 0 \  95\%$ C.L.}}
\multiput(1328,632)(20.756,0.000){4}{\usebox{\plotpoint}}
\put(1394,632){\usebox{\plotpoint}}
\put(420,832){\usebox{\plotpoint}}
\multiput(420,832)(0.000,-20.756){9}{\usebox{\plotpoint}}
\multiput(420,652)(0.116,-20.755){9}{\usebox{\plotpoint}}
\multiput(421,473)(1.209,-20.720){6}{\usebox{\plotpoint}}
\multiput(428,353)(3.743,-20.415){3}{\usebox{\plotpoint}}
\put(447.02,274.49){\usebox{\plotpoint}}
\multiput(452,263)(13.885,-15.427){2}{\usebox{\plotpoint}}
\put(487.05,228.17){\usebox{\plotpoint}}
\multiput(504,218)(19.434,-7.288){2}{\usebox{\plotpoint}}
\put(544.28,204.62){\usebox{\plotpoint}}
\put(564.50,200.00){\usebox{\plotpoint}}
\put(585.02,197.00){\usebox{\plotpoint}}
\put(605.63,194.58){\usebox{\plotpoint}}
\multiput(625,192)(20.720,-1.219){5}{\usebox{\plotpoint}}
\multiput(727,186)(20.753,-0.341){15}{\usebox{\plotpoint}}
\multiput(1031,181)(20.756,0.000){20}{\usebox{\plotpoint}}
\put(1436,181){\usebox{\plotpoint}}
\end{picture}

\end{document}